\documentclass[twocolumn,superscriptaddress,floatfix,preprintnumbers]{revtex4}

\usepackage{graphics}
\usepackage{amsmath}
\usepackage{amsfonts}
\usepackage{amssymb}
\usepackage{graphicx}

\begin{document}
\title{Inverse design and demonstration of a compact on-chip narrowband three-channel wavelength demultiplexer}

\author{Logan Su}
\author{Alexander Y. Piggott}
\author{Neil V. Sapra}
\author{Jan Petykiewicz}
\author{Jelena Vu\v{c}kovi\'{c}}
\affiliation{Ginzton Laboratory, Stanford University, Stanford, California 94305, USA}

\begin{abstract}
In wavelength division multiplexing (WDM) schemes, splitters must be used to combine and separate different wavelengths. Conventional splitters are fairly large with footprints in hundreds to thousands of square microns, and experimentally-demonstrated MMI-based and inverse-designed ultra-compact splitters operate with only two channels and large channel spacing ($>$100 nm). Here we inverse design and experimentally demonstrate a three-channel wavelength demultiplexer with 40 nm spacing (1500 nm, 1540 nm, and 1580 nm) with a footprint of 24.75 $\mu\mathrm{m}^2$. The splitter has a simulated peak insertion loss of -1.55 dB with under -15 dB crosstalk and a measured peak insertion loss of -2.29 dB with under -10.7 dB crosstalk.
\end{abstract}

\maketitle

\section{Introduction}
Integrated silicon photonics can play key roles in many applications, including optical interconnects \cite{Miller:2017} and quantum technologies \cite{Kok:2007}. One of the advantages to using photonics is utilizing different wavelengths of light to carry information in order to dramatically increase the information bandwidth in a fiber or waveguide. In such wavelength division multiplexing (WDM) systems, wavelength demultiplexers are used to separate the different channels. Conventional demultiplexers, such as ring resonator arrays and arrayed waveguide gratings, have fairly large footprints \cite{Bogaerts:2010, Dai:2017}. Nanophotonic inverse design has enabled the design of more compact devices \cite{Lu:13, Keraly:13, Piggott:14, Shen:15, Frellsen:16}, but previous experimental demonstrations of inverse-designed wavelength demultiplexers have only achieved broadband demultiplexing of two channels with large channel spacings ($>$100 nm) \cite{Piggott:15, Borel:07}. Frandsen et al. experimentally showed a  drop-filter with 11 nm full-width-half-maximum (FWHM) \cite{Frandsen:2013}, but the device only filters out one wavelength over a larger footprint than we demonstrate here. Recent experimental demonstrations of multimode interference (MMI) devices have also achieved demultiplexing capabilities in ultra-compact footprints but again with large channel spacings over two channels \cite{Mu:16, Rouifed:17}. Since the number of wavelengths available in a WDM system is inversely proportional to the channel spacing, WDM systems that utilize multiple wavelengths require demultiplexers with more channels and much smaller channel spacing. Here, using our nanophotonic inverse design approach \cite{Lu:13, Piggott:15, Piggott:17}, we design and experimentally demonstrate a three-channel wavelength demultiplexer with 40 nm channel spacing and a 5.5 $\mu$m $\times$ 4.5 $\mu$m footprint for the silicon-on-insulator (SOI) platform.

\section{Design Method}

The setup of the problem follows closely with our work described in the references \cite{Piggott:15} and \cite{Piggott:17}. For clarity, we provide a brief overview and highlight the salient points of the optimization in Section \ref{sec:setup}. Further details can be found in the supplementary information of \cite{Piggott:17}. In Section \ref{sec:biasing}, we discuss modifications to the optimization process used to design the narrow-band demultiplexer. In Section \ref{sec:design}, we describe the complete design algorithm with the specific parameters used in the optimization and design.

\subsection{Setup}
\label{sec:setup}

The design of a 3-channel wavelength demultiplexer is inherently a multiobjective problem: At each operating wavelength, it is desirable to maximize the transmission  while at the same time minimizing the crosstalk. The relative importance of these goals are expressed through an objective function $F$. The exact form of $F$ is detailed in the supplementary information of \cite{Piggott:17}. 

Applying the fabrication-constrained optimization procedure outlined in \cite{Piggott:17} requires starting with a good initial condition as the optimization landscape is highly non-convex with many undesirable local optima. Consequently the optimization procedure is split into two stages. In the first stage, deemed {\it continuous optimization}, the discrete constraint is relaxed to allow the permittivities to vary continuously between that of silicon oxide and silicon. This optimization stage provides a structure that seeds the second stage, fabrication-constrained {\it discrete optimization}, which, as its name implies, produces a fabricable, discrete structure.

In the continuous stage, the structure is parametrized by a 2D image where each pixel of the image corresponds to the permittivity of the device at the corresponding location. A local optimum of the objective can be found by applying gradient descent. Naively calculating the gradient is an expensive operation since it involves computing the variation of the objective with every design component. However, using the adjoint method \cite{Lee:1997}, the gradient calculation is related to the result of a time-reversed electromagnetic simulation. Consequently, the gradient can be computed by running a single simulation, which is substantially more efficient than the naive approach.

In the discrete stage, the device is parametrized by a spatially continuous level set function, where the permittivity of the device at a particular location depends on the sign of the level set function. The level set representation can characterize arbitrary boundaries and naturally handles merging and splitting of holes. A gradient-descent-like update can be performed on the level set and can be extended to impose fabrication constraints \cite{Piggott:17}.

\subsection{Biasing}
\label{sec:biasing}

In its simplest form, the continuous optimization stage does not always generate a good initial condition for the discrete stage. Empirically, we found that the outcome of the continuous optimization stage provides a good initial condition for the fabrication-constrained discrete optimization stage if the output structure of the continuous optimization is nearly discrete, i.e. each pixel has a permittivity close to that of the device or that of the cladding. Unfortunately, applying gradient descent directly can lead to structures that have weakly-modulated permittivities that are somewhere in between, and this results in poorly performing structures in the discrete stage. It is possible to optimize in the continuous stage for more steps in the hopes of obtaining a more discrete structure, but in practice, this not only requires a significant overhead in computational resources but also offers no guarantee that a discrete structure will eventually form.

There are a wide variety of proposals to mitigate this issue of ``gray'' areas in the structure, including density filters \cite{Bruns:01}, sensitivity filters \cite{Sigmund:97}, penalty functions \cite{Allaire:93}, artificial damping \cite{Sigmund:05}, and morphological filters \cite{Sigmund:07}. In this work, we mitigate this issue through a specific variant of penalty functions, which we call biasing. Specifically, we introduce the concept of self-biasing to produce more discrete pixels and the concept of neighbor biasing to produce discrete structures with larger feature sizes.

As noted previously, in the continuous stage, the structure is parametrized by a 2D image. More specifically, the structure can be described by a vector $\mathbf{z} \in [0,1]^{n}$ where $n$ is the total number of pixels in the image. The permittivity at the $i$th pixel is given by $\epsilon_i = (\epsilon_{hi} - \epsilon_{lo})z_i + \epsilon_{lo}$ where $\epsilon_{hi}$ is the permittivity of the device and $\epsilon_{lo}$ is that of the cladding. These are silicon and silicon oxide, respectively, for the wavelength demultiplexer.

The gradient descent update can be described by the operation
\begin{equation}
	\mathbf{z} \gets \mathbf{z} - \alpha \nabla F(\mathbf{z})
\end{equation}
where $\alpha$ is the gradient descent step size and $F$ is the objective function. After the gradient update we perform a {\it self-biasing update} to the parametrization:
\begin{equation}
	\mathbf{z} \gets \text{clip}(b_s(\mathbf{z}))
\end{equation}
where the clipping function, applied element-wise to the vector, is defined as
\begin{equation}
	\text{clip}(x) = \begin{cases}
x,& \text{if } 0 \leq x \leq 1 \\
0,              & \text{if } x < 0 \\
1&  \text{if } x > 1
\end{cases}
\end{equation}
and the self-biasing function $b_s$ is defined as 
\begin{equation}
	b_s(\mathbf{z}) = \frac{1}{1 - 2k}\left(\mathbf{z} - \frac{1}{2}\right) + \frac{1}{2}
\end{equation}
and $0 \leq k < \frac{1}{2}$ is a parameter to the biasing function. The goal of the biasing function is shift the pixel values towards either 0 or 1, corresponding to $\epsilon_{lo}$ and $\epsilon_{hi}$, respectively. Indeed, $b_s(z_i) > z_i$ when $z_i > \frac{1}{2}$ and $b_s(z_i) < z_i$ when $z_i < \frac{1}{2}$. The value of $k$ determines the strength of this biasing: a large $k$ corresponds to a dramatic change in $z_i$ whereas a small $k$ corresponds to a gentle push in $z_i$. The name self-biasing comes from the fact that each pixel is biased towards zero or one depending on its current value.

By combining the gradient descent update and the self-biasing update into one update step, it is not hard to show that the self-biasing update is equivalent to adding a quadratic penalty function $g$ to the objective of the form $g(\mathbf{z}) = c\mathbf{z}^T(\mathbf{1}-\mathbf{z})$ for some constant $c$. However, there are several advantages to expressing self-biasing as a separate update step. First, the discretization goal often opposes progress towards a well-performing device. A line search is often used in gradient-based methods in order to speed up (and ensure) convergence. Under a line search, the objective function is forced to decrease in value each iteration, and empirically, incorporating self-biasing update into the objective function results in premature convergence to poor solutions. Second, expressing self-biasing as a separate update readily generalizes to more sophisticated types of biasing, as we will see shortly.

\begin{figure}[htbp]
	\centering
	\includegraphics[width=\linewidth]{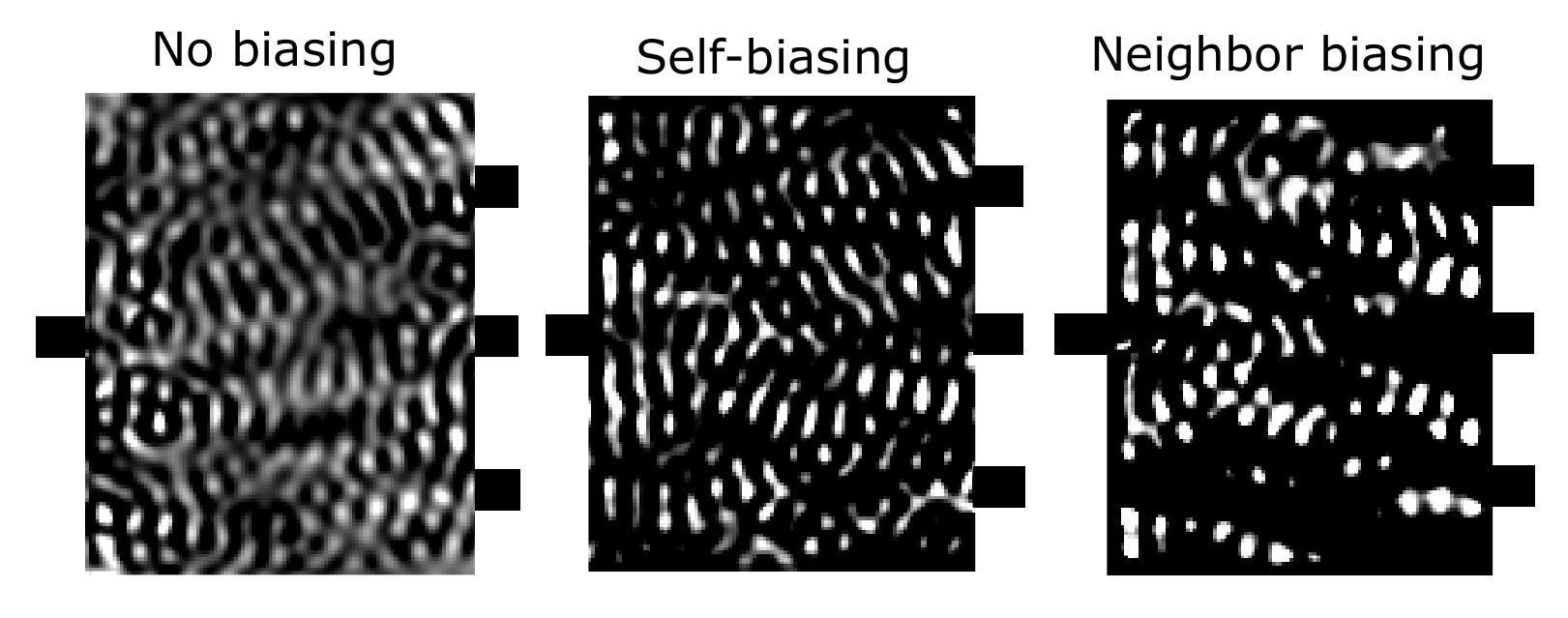}
	\caption{Optimized continuous structures with no biasing, self-biasing, and neighbor biasing after 100 iterations. Black represents silicon (1) and white represents silica (0). With no biasing, the structure is not discrete at all. With self-biasing, the structure becomes much more discrete but at the expense of producing many small features. With neighbor biasing, the structure becomes discrete and avoids smaller features.}
	\label{fig:biasing}
\end{figure}

In Figure \ref{fig:biasing}, we see the results with and without self-biasing applied. As one can see, self-biasing often produces nearly-discrete structures that have very small features. The reason is that self-biasing is a self-reinforcing action: A pixel that was biased towards zero/one in one iteration will likely be biased towards zero/one in the next iteration. In order to address this issue, we introduce the notion of {\it neighbor biasing}, in which the pixel values are biased based not only on their current values but the values of their neighbors as well. Mathematically, we choose to use the neighbor biasing function $b_n$ in place of $b_s$ where
\begin{equation}
	b_n(\mathbf{z}) = \frac{1}{1 - 2k}(\mathbf{z} - h(\mathbf{z})) + \frac{1}{2}
\end{equation}
and $h$ is a function whose $i$th component arises from averaging the pixel values in a circle of radius $r$ around the $i$th pixel. The averaging radius $r$ controls the size of the holes in the structure.

To compute $h(\mathbf{z})$, we first treat $\mathbf{z}$ as a 2D grayscale image. We convolve this image with a uniform circular disk of radius $r$, and denote the resulting image as $\mathbf{z}_{avg}$. We then define $h(\mathbf{z})$ as
\begin{equation}
[h(\mathbf{z})]_i = \frac{1}{2} + k_{avg} \left(\frac{1}{2} - [z_{avg}]_i\right)^{2p+1}
\end{equation}
where $[v]_i$ denotes the $i$th element of vector $v$, and $k_{avg}$ and $p$ are parameters that control the strength of the bias. The parameter $p$ reflects the fact that biasing should be weak if $[z_{avg}]_i$ is close to $\frac{1}{2}$ but substantially stronger if $[z_{avg}]_i$ is far from $\frac{1}{2}$. The parameter $k_{avg}$ scales the biasing depending on the chosen $p$ in order to not discretize the structure too quickly. Notice that the neighbor-biasing update is equivalent to self-biasing update when $[z_{avg}]_i = \frac{1}{2}$. Moreover, note that in this biasing scheme, it is easier mathematically and more intuitive to work with the update step directly rather than expressing the result as a penalty function.

Figure \ref{fig:biasing} compares the results of using no-biasing, self-biasing, and neighbor biasing. Under no biasing, there a large regions of intermediate permittivity. With self-biasing, these intermediate regions largely disappear but at the expense of small features, but with neighbor biasing, the structure is both discrete and mostly free of small features.

\subsection{Design of 3-channel Wavelength Demultiplexer}
\label{sec:design}
The 3-channel wavelength demultiplexer is designed on single fully-etched 220 nm thick Si layer with $\mathrm{SiO}_2$ cladding. Refractive indices of $n_\text{Si} = 3.48$ and $n_{\text{SiO}_2}$ = 1.44 were used. The waveguide width was set to 500 nm for both the input and output waveguides. The demultiplexer was designed for operation at 1500 nm, 1540 nm, and 1580 nm. The power in the fundamental transverse-electric (TE) mode of the input waveguide at each wavelength was maximized at the corresponding output waveguide and minimized at the other two waveguides.
 
 We solved the optimization problem given by
 \begin{equation}
 \begin{aligned}
 & \underset{\mathbf{E}_1, \mathbf{E}_2, \mathbf{E}_3, \mathbf{\phi}}{\text{minimize}}
 & & F(\mathbf{E}_1, \mathbf{E}_2, \mathbf{E}_3) \\
 & \text{subject to}
 & & \nabla \times \frac{1}{\mu_0} \nabla \times \mathbf{E}_i - \omega^2 \epsilon(\mathbf{\phi})\mathbf{E}_i   = -i\omega\mathbf{J}_i, \\
 & & & i = 1, 2, 3
 \end{aligned}
 \end{equation}
 where $\mathbf{E}_i$ is the electric field at the $i$th wavelength (i.e. 1500 nm, 1540 nm, or 1580 nm), $\mathbf{J}_i$ is the total-field scattered-field (TFSF) current source to excite the TE mode of the input waveguide, and $\mathbf{\phi}$ parametrizes the structure. The form of $\phi$ depends on the stage of the optimization (continuous or discrete).
 The objective function $F$ is given by
 \begin{align*}
 F(\mathbf{E}_1, \mathbf{E}_2, \mathbf{E}_3) = \sum_{i=1}^3 f_i(\mathbf{E}_i)
 \end{align*}
where $f_i$ represents the sub-objective for the $i$th wavelength and is given by equation S16 in \cite{Piggott:17}:
 \begin{equation}
 f_i(\mathbf{x}_i) = \sum_{j=1}^3 I_+\left(\left|\mathcal{L}(\mathbf{E}_i)\right| - \alpha_{ij}\right) + I_-\left(\beta_{ij} - \left|\mathcal{L}(\mathbf{E}_i)\right|\right)
 \end{equation}
 where $I_+$ and $I_-$ are relaxed indicator functions as defined in equation S17 in \cite{Piggott:17}.  Here, $\alpha_{ij}$ and $\beta_{ij}$ are chosen to be 0.99 and 1, respectively, when $i = j$ and chosen to be 0 and 0.01, respectively, when $i \neq j$. $\mathcal{L}(\mathbf{E}_i)$ is given by
 \begin{equation}
	\mathcal{L}(\mathbf{E}_i) = \iint \left(\mathbf{E}_i \times H_{ij} + E_{ij} \times \frac{i}{\omega\mu_0}  \nabla \times \mathbf{E}_i\right) \cdot \hat{n} d\mathbf{r}_\perp	
 \end{equation}
 where $E_{ij}$ and $H_{ij}$ are the electric and magnetic fields of the $j$th output mode at the $i$th wavelength, $\hat{n}$ is the normalized vector in the waveguide propagation direction, and $\mathbf{r}_\perp$ represents the coordinates orthogonal to the propagation direction.
 The supplementary material for \cite{Piggott:17} contains the full details on the definitions and mathematical derivation of the gradients.

The structure was first optimized in the continuous stage wherein $\mathbf{\epsilon} = \mathbf{\phi}$ for 210 iterations. During this stage, neighbor biasing with $k = 0.01$, $k_{avg} = 0.2$ and $p = 3$ was employed. The resulting structure was thresholded at a threshold level of 0.5 and used as the initial condition for the discrete stage optimization. The discrete stage optimization ran for 145 iterations and follows the procedure identical to our prior work \cite{Piggott:17}. The minimum radius of curvature was constrained to be 40 nm and minimum width of a hole to be 90 nm.

The device was designed in approximately 60 hours on a single computer with an Intel Core i7-5820K processor,
64GB of RAM, and three Nvidia Titan Z graphics cards. All electromagnetic simulations were performed using a graphical
processing unit (GPU) accelerated implementation of the finite-difference frequency-domain (FDFD) method \cite{Shin:12, Shin:13} with a spatial step size of 40 nm.

\section{Experimental Results}

\begin{figure*}[htbp]
	\centering
	\includegraphics[scale=1]{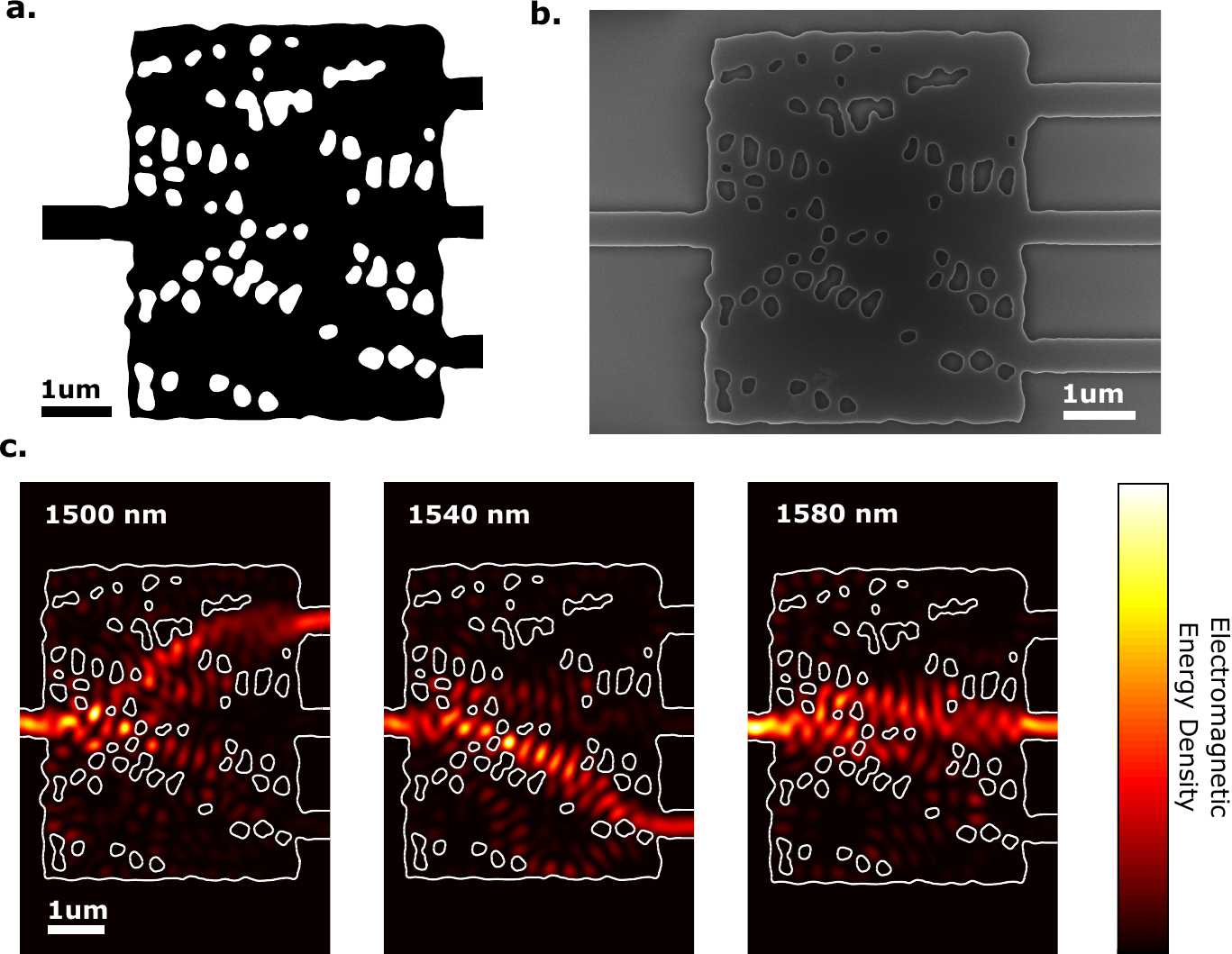}
	\caption{The three-channel wavelength demultiplexer. (a) Design of the device. Black represents silicon and white is silica. (b) SEM image of the fabricated device. The total footprint is 5.5 $\mu$m $\times$ 4.5 $\mu$m. (c) Simulated electromagnetic energy density ($U = \frac{1}{2}\epsilon|\mathbf{E}|^2 + \frac{1}{2}\mu|\mathbf{H}|^2$) in the device at the three operating wavelengths.}
	\label{fig:device}
\end{figure*}

\subsection{Fabrication}
The wavelength demultiplexer was fabricated on Unibond SmartCut silicon-on-insulator (SOI) wafers obtained from SOITEC, with a nominal 220 nm device layer and 3.0 $\mu$m buried oxide layer. A JEOL JBX-6300FS electron-beam lithography system was used to pattern a 330 nm thick layer of ZEP-520A resist spun on the samples. No proximity-effect correction step was performed. A transformer-coupled plasma etcher was used to transfer the pattern to the device layer, using a $\mathrm{C}_2\mathrm{F}_6$ breakthrough
step and HBr/$\mathrm{O}_2$/He main etch. The mask was stripped by soaking in solvents, followed by a HF dip. Finally, the devices were capped with 1.6 $\mu$m of low pressure chemical vapor deposition (LPCVD) oxide.

A multi-step etch-based process was used to expose waveguide facets for edge coupling. First, a chrome mask was deposited using liftoff to protect the devices. Next, the oxide cladding, device layer, and buried oxide layer were etched in a inductively-coupled plasma etcher using a $\mathrm{C}_4\mathrm{F}_8$/$\mathrm{O}_2$ chemistry. Next, a protective 20 nm $\mathrm{Al}_2\mathrm{O}_3$ coating was deposited using atomic layer deposition (ALD) in order to protect the waveguide facets during further processing. An anisotropic etch using a $\mathrm{Cl}_2$/$\mathrm{BCl}_3$/$\mathrm{N}_2$ chemistry removed the $\mathrm{Al}_2\mathrm{O}_3$ coating on the substrate. To provide mechanical clearance for the optical fibers, the silicon substrate was etched to a depth of around 100 $\mu$m using the Bosch process in a deep reactive-ion etcher (DRIE). Finally, the chrome mask was chemically stripped, and the samples were diced into conveniently-sized pieces.

\subsection{Characterization}

\begin{figure*}[htb]
	\centering
	\includegraphics[scale=0.6]{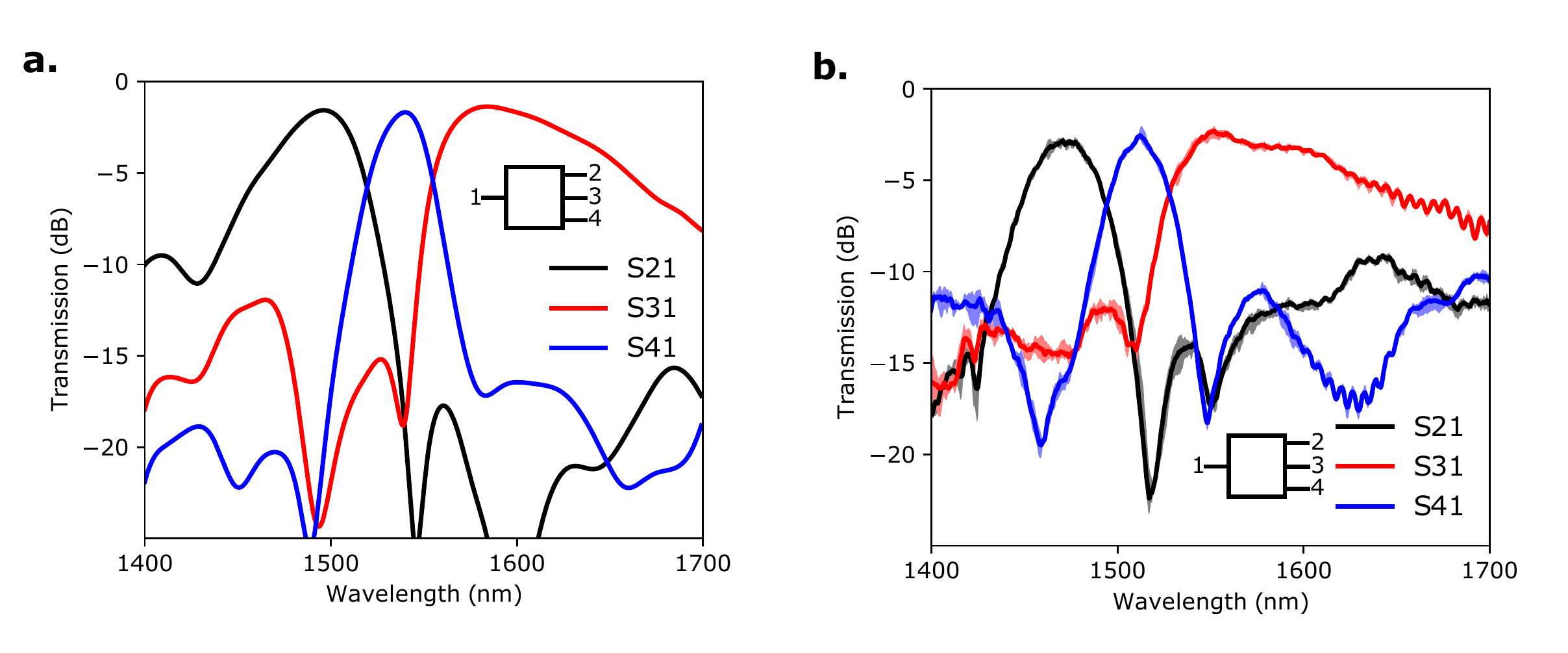}
	\caption{Simulated and measured S-parameters for the WDM, where $S_{ij}$ is the transmission from port $j$ to port $i$. (a) Simulated transmission calculated using finite-time finite-difference (FDTD). (b) Measured transmission of four identically fabricated devices. The solid lines indicate the average of the four devices, and the shaded region is bounded by the minimum and maximum measured transmission.}
	\label{fig:perf}
\end{figure*}

Figure \ref{fig:device} shows the design, scanning electron microscope (SEM) image of the fabricated device, and simulated electromagnetic density at the operating wavelengths.

The transmission through the device was measured by edge-coupling input and output waveguides with lensed fibers. A polarization-maintaining fiber was used at the input to ensure that only the TE mode of the silicon waveguide was excited. Consistent edge-coupling was achieved by maximizing transmission with a 1570 nm laser, and the transmission spectra were measured with a supercontinuum source and spectrum analyzer. The spectra were normalized against transmission through a waveguide adjacent to the device.

The simulated and measured transmission are shown in Figure \ref{fig:perf}. The consistency of the measurements across four identically fabricated devices indicate that the device is robust to fabrication imprecision. The peak simulated transmission was -1.56 dB at 1500 nm, -1.68 dB at 1540 nm, and -1.35 dB at 1580 nm. The peak average measured transmission was -2.82 dB at 1471 nm, -2.55 at 1512 nm, and -2.29 dB at 1551 nm. At peak transmission, simulated crosstalk was under -15 dB, and measured crosstalk was under -10.7 dB. The discrepancies between simulated and measured devices are a likely result of slight underetching and/or overetching during the fabrication process.

\section{Conclusion}

By using a biasing technique in the optimization process, we have designed and experimentally demonstrated an efficient, compact, narrowband three-channel wavelength demultiplexer on SOI. The consistent performance across four fabricated devices indicate that the designs are also robust to fabrication errors. We expect that similar inverse design techniques can be used to design demultiplexers with more channels and smaller channel spacing while maintaining a relatively small footprint as compared to conventional demultiplexers.

\section*{Acknowledgements}
This work was funded by the AFOSR MURI for Aperiodic Silicon Photonics, grant number FA9550-15-1-0335,
the Gordon and Betty Moore Foundation, and GlobalFoundries Inc. All devices were fabricated at the Stanford
Nanofabrication Facility (SNF) and Stanford Nano Shared Facilities (SNSF).


\begin{thebibliography}{23}
	\expandafter\ifx\csname natexlab\endcsname\relax\def\natexlab#1{#1}\fi
	\expandafter\ifx\csname bibnamefont\endcsname\relax
	\def\bibnamefont#1{#1}\fi
	\expandafter\ifx\csname bibfnamefont\endcsname\relax
	\def\bibfnamefont#1{#1}\fi
	\expandafter\ifx\csname citenamefont\endcsname\relax
	\def\citenamefont#1{#1}\fi
	\expandafter\ifx\csname url\endcsname\relax
	\def\url#1{\texttt{#1}}\fi
	\expandafter\ifx\csname urlprefix\endcsname\relax\def\urlprefix{URL }\fi
	\providecommand{\bibinfo}[2]{#2}
	\providecommand{\eprint}[2][]{\url{#2}}
	
	\bibitem[{\citenamefont{Miller}(2017)}]{Miller:2017}
	\bibinfo{author}{\bibfnamefont{D.~A.} \bibnamefont{Miller}},
	\bibinfo{journal}{Journal of Lightwave Technology}
	\textbf{\bibinfo{volume}{35}}, \bibinfo{pages}{346} (\bibinfo{year}{2017}).
	
	\bibitem[{\citenamefont{Kok et~al.}(2007)\citenamefont{Kok, Munro, Nemoto,
			Ralph, Dowling, and Milburn}}]{Kok:2007}
	\bibinfo{author}{\bibfnamefont{P.}~\bibnamefont{Kok}},
	\bibinfo{author}{\bibfnamefont{W.~J.} \bibnamefont{Munro}},
	\bibinfo{author}{\bibfnamefont{K.}~\bibnamefont{Nemoto}},
	\bibinfo{author}{\bibfnamefont{T.~C.} \bibnamefont{Ralph}},
	\bibinfo{author}{\bibfnamefont{J.~P.} \bibnamefont{Dowling}},
	\bibnamefont{and} \bibinfo{author}{\bibfnamefont{G.~J.}
		\bibnamefont{Milburn}}, \bibinfo{journal}{Reviews of Modern Physics}
	\textbf{\bibinfo{volume}{79}}, \bibinfo{pages}{135} (\bibinfo{year}{2007}).
	
	\bibitem[{\citenamefont{Bogaerts et~al.}(2010)\citenamefont{Bogaerts,
			Selvaraja, Dumon, Brouckaert, Vos, Thourhout, and Baets}}]{Bogaerts:2010}
	\bibinfo{author}{\bibfnamefont{W.}~\bibnamefont{Bogaerts}},
	\bibinfo{author}{\bibfnamefont{S.~K.} \bibnamefont{Selvaraja}},
	\bibinfo{author}{\bibfnamefont{P.}~\bibnamefont{Dumon}},
	\bibinfo{author}{\bibfnamefont{J.}~\bibnamefont{Brouckaert}},
	\bibinfo{author}{\bibfnamefont{K.~D.} \bibnamefont{Vos}},
	\bibinfo{author}{\bibfnamefont{D.~V.} \bibnamefont{Thourhout}},
	\bibnamefont{and} \bibinfo{author}{\bibfnamefont{R.}~\bibnamefont{Baets}},
	\bibinfo{journal}{IEEE Journal of Selected Topics in Quantum Electronics}
	\textbf{\bibinfo{volume}{16}}, \bibinfo{pages}{33} (\bibinfo{year}{2010}).
	
	\bibitem[{\citenamefont{Dai}(2017)}]{Dai:2017}
	\bibinfo{author}{\bibfnamefont{D.}~\bibnamefont{Dai}},
	\bibinfo{journal}{Journal of Lightwave Technology}
	\textbf{\bibinfo{volume}{35}}, \bibinfo{pages}{572} (\bibinfo{year}{2017}).
	
	\bibitem[{\citenamefont{Lu and Vu\v{c}kovi\'{c}}(2013)}]{Lu:13}
	\bibinfo{author}{\bibfnamefont{J.}~\bibnamefont{Lu}} \bibnamefont{and}
	\bibinfo{author}{\bibfnamefont{J.}~\bibnamefont{Vu\v{c}kovi\'{c}}},
	\bibinfo{journal}{Optics Express} \textbf{\bibinfo{volume}{21}},
	\bibinfo{pages}{13351} (\bibinfo{year}{2013}).
	
	\bibitem[{\citenamefont{Lalau-Keraly et~al.}(2013)\citenamefont{Lalau-Keraly,
			Bhargava, Miller, and Yablonovitch}}]{Keraly:13}
	\bibinfo{author}{\bibfnamefont{C.~M.} \bibnamefont{Lalau-Keraly}},
	\bibinfo{author}{\bibfnamefont{S.}~\bibnamefont{Bhargava}},
	\bibinfo{author}{\bibfnamefont{O.~D.} \bibnamefont{Miller}},
	\bibnamefont{and}
	\bibinfo{author}{\bibfnamefont{E.}~\bibnamefont{Yablonovitch}},
	\bibinfo{journal}{Optics Express} \textbf{\bibinfo{volume}{21}},
	\bibinfo{pages}{21693} (\bibinfo{year}{2013}).
	
	\bibitem[{\citenamefont{Piggott et~al.}(2014)\citenamefont{Piggott, Lu,
			Babinec, Lagoudakis, Petykiewicz, and Vu\v{c}kovi\'{c}}}]{Piggott:14}
	\bibinfo{author}{\bibfnamefont{A.~Y.} \bibnamefont{Piggott}},
	\bibinfo{author}{\bibfnamefont{J.}~\bibnamefont{Lu}},
	\bibinfo{author}{\bibfnamefont{T.~M.} \bibnamefont{Babinec}},
	\bibinfo{author}{\bibfnamefont{K.~G.} \bibnamefont{Lagoudakis}},
	\bibinfo{author}{\bibfnamefont{J.}~\bibnamefont{Petykiewicz}},
	\bibnamefont{and}
	\bibinfo{author}{\bibfnamefont{J.}~\bibnamefont{Vu\v{c}kovi\'{c}}},
	\bibinfo{journal}{Scientific Reports} \textbf{\bibinfo{volume}{4}}
	(\bibinfo{year}{2014}).
	
	\bibitem[{\citenamefont{Shen et~al.}(2015)\citenamefont{Shen, Wang, Polson, and
			Menon}}]{Shen:15}
	\bibinfo{author}{\bibfnamefont{B.}~\bibnamefont{Shen}},
	\bibinfo{author}{\bibfnamefont{P.}~\bibnamefont{Wang}},
	\bibinfo{author}{\bibfnamefont{R.}~\bibnamefont{Polson}}, \bibnamefont{and}
	\bibinfo{author}{\bibfnamefont{R.}~\bibnamefont{Menon}},
	\bibinfo{journal}{Nature Photonics} \textbf{\bibinfo{volume}{9}},
	\bibinfo{pages}{378} (\bibinfo{year}{2015}).
	
	\bibitem[{\citenamefont{Frellsen et~al.}(2016)\citenamefont{Frellsen, Ding,
			Sigmund, and Frandsen}}]{Frellsen:16}
	\bibinfo{author}{\bibfnamefont{L.~F.} \bibnamefont{Frellsen}},
	\bibinfo{author}{\bibfnamefont{Y.}~\bibnamefont{Ding}},
	\bibinfo{author}{\bibfnamefont{O.}~\bibnamefont{Sigmund}}, \bibnamefont{and}
	\bibinfo{author}{\bibfnamefont{L.~H.} \bibnamefont{Frandsen}},
	\bibinfo{journal}{Optics Express} \textbf{\bibinfo{volume}{24}},
	\bibinfo{pages}{16866} (\bibinfo{year}{2016}).
	
	\bibitem[{\citenamefont{Piggott et~al.}(2015)\citenamefont{Piggott, Lu,
			Lagoudakis, Petykiewicz, Babinec, and Vu\v{c}kovi\'{c}}}]{Piggott:15}
	\bibinfo{author}{\bibfnamefont{A.~Y.} \bibnamefont{Piggott}},
	\bibinfo{author}{\bibfnamefont{J.}~\bibnamefont{Lu}},
	\bibinfo{author}{\bibfnamefont{K.~G.} \bibnamefont{Lagoudakis}},
	\bibinfo{author}{\bibfnamefont{J.}~\bibnamefont{Petykiewicz}},
	\bibinfo{author}{\bibfnamefont{T.~M.} \bibnamefont{Babinec}},
	\bibnamefont{and}
	\bibinfo{author}{\bibfnamefont{J.}~\bibnamefont{Vu\v{c}kovi\'{c}}},
	\bibinfo{journal}{Nature Photonics} \textbf{\bibinfo{volume}{9}},
	\bibinfo{pages}{374} (\bibinfo{year}{2015}).
	
	\bibitem[{\citenamefont{Borel et~al.}(2007)\citenamefont{Borel, Bilenberg,
			Frandsen, Nielsen, Fage-Pedersen, Lavrinenko, Jensen, Sigmund, , and
			Kristensen}}]{Borel:07}
	\bibinfo{author}{\bibfnamefont{P.~I.} \bibnamefont{Borel}},
	\bibinfo{author}{\bibfnamefont{B.}~\bibnamefont{Bilenberg}},
	\bibinfo{author}{\bibfnamefont{L.~H.} \bibnamefont{Frandsen}},
	\bibinfo{author}{\bibfnamefont{T.}~\bibnamefont{Nielsen}},
	\bibinfo{author}{\bibfnamefont{J.}~\bibnamefont{Fage-Pedersen}},
	\bibinfo{author}{\bibfnamefont{A.~V.} \bibnamefont{Lavrinenko}},
	\bibinfo{author}{\bibfnamefont{J.~S.} \bibnamefont{Jensen}},
	\bibinfo{author}{\bibfnamefont{O.}~\bibnamefont{Sigmund}}, ,
	\bibnamefont{and}
	\bibinfo{author}{\bibfnamefont{A.}~\bibnamefont{Kristensen}},
	\bibinfo{journal}{Optics Express} \textbf{\bibinfo{volume}{15}},
	\bibinfo{pages}{1261} (\bibinfo{year}{2007}).
	
	\bibitem[{\citenamefont{Frandsen et~al.}(2013)\citenamefont{Frandsen, Elesin,
			Sigmund, Jensen, and Yvind}}]{Frandsen:2013}
	\bibinfo{author}{\bibfnamefont{L.~H.} \bibnamefont{Frandsen}},
	\bibinfo{author}{\bibfnamefont{Y.}~\bibnamefont{Elesin}},
	\bibinfo{author}{\bibfnamefont{O.}~\bibnamefont{Sigmund}},
	\bibinfo{author}{\bibfnamefont{J.~S.} \bibnamefont{Jensen}},
	\bibnamefont{and} \bibinfo{author}{\bibfnamefont{K.}~\bibnamefont{Yvind}}, in
	\emph{\bibinfo{booktitle}{CLEO: Science and Innovations}}
	(\bibinfo{organization}{Optical Society of America}, \bibinfo{year}{2013}),
	pp. \bibinfo{pages}{CTh4L--6}.
	
	\bibitem[{\citenamefont{Mu et~al.}(2016)\citenamefont{Mu, Vázquez-Córdova,
			Sefunc, Yong, and García-Blanco}}]{Mu:16}
	\bibinfo{author}{\bibfnamefont{J.}~\bibnamefont{Mu}},
	\bibinfo{author}{\bibfnamefont{S.~A.} \bibnamefont{Vázquez-Córdova}},
	\bibinfo{author}{\bibfnamefont{M.~A.} \bibnamefont{Sefunc}},
	\bibinfo{author}{\bibfnamefont{Y.-S.} \bibnamefont{Yong}}, \bibnamefont{and}
	\bibinfo{author}{\bibfnamefont{S.~M.} \bibnamefont{García-Blanco}},
	\bibinfo{journal}{Journal of Lightwave Technology}
	\textbf{\bibinfo{volume}{34}}, \bibinfo{pages}{3603} (\bibinfo{year}{2016}).
	
	\bibitem[{\citenamefont{Rouifed et~al.}(2017)\citenamefont{Rouifed,
			Littlejohns, Tina, Qiu, Penades, Nedeljkovic, Zhang, Liu, Thomson,
			Mashanovich et~al.}}]{Rouifed:17}
	\bibinfo{author}{\bibfnamefont{M.-S.} \bibnamefont{Rouifed}},
	\bibinfo{author}{\bibfnamefont{C.~G.} \bibnamefont{Littlejohns}},
	\bibinfo{author}{\bibfnamefont{G.~X.} \bibnamefont{Tina}},
	\bibinfo{author}{\bibfnamefont{H.}~\bibnamefont{Qiu}},
	\bibinfo{author}{\bibfnamefont{J.~S.} \bibnamefont{Penades}},
	\bibinfo{author}{\bibfnamefont{M.}~\bibnamefont{Nedeljkovic}},
	\bibinfo{author}{\bibfnamefont{Z.}~\bibnamefont{Zhang}},
	\bibinfo{author}{\bibfnamefont{C.}~\bibnamefont{Liu}},
	\bibinfo{author}{\bibfnamefont{D.~J.} \bibnamefont{Thomson}},
	\bibinfo{author}{\bibfnamefont{G.~Z.} \bibnamefont{Mashanovich}},
	\bibnamefont{et~al.}, \bibinfo{journal}{Optics Express}
	\textbf{\bibinfo{volume}{25}}, \bibinfo{pages}{10893} (\bibinfo{year}{2017}).
	
	\bibitem[{\citenamefont{Piggott et~al.}(2017)\citenamefont{Piggott,
			Petykiewicz, Su, and Vu\v{c}kovi\'{c}}}]{Piggott:17}
	\bibinfo{author}{\bibfnamefont{A.~Y.} \bibnamefont{Piggott}},
	\bibinfo{author}{\bibfnamefont{J.}~\bibnamefont{Petykiewicz}},
	\bibinfo{author}{\bibfnamefont{L.}~\bibnamefont{Su}}, \bibnamefont{and}
	\bibinfo{author}{\bibfnamefont{J.}~\bibnamefont{Vu\v{c}kovi\'{c}}},
	\bibinfo{journal}{Scientific Reports} \textbf{\bibinfo{volume}{7}},
	\bibinfo{pages}{1786} (\bibinfo{year}{2017}).
	
	\bibitem[{\citenamefont{Lee and Itoh}(1997)}]{Lee:1997}
	\bibinfo{author}{\bibfnamefont{H.-b.} \bibnamefont{Lee}} \bibnamefont{and}
	\bibinfo{author}{\bibfnamefont{T.}~\bibnamefont{Itoh}},
	\bibinfo{journal}{IEEE Transactions on Microwave Theory and Techniques}
	\textbf{\bibinfo{volume}{45}}, \bibinfo{pages}{803} (\bibinfo{year}{1997}).
	
	\bibitem[{\citenamefont{Bruns and Tortorelli}(2001)}]{Bruns:01}
	\bibinfo{author}{\bibfnamefont{T.~E.} \bibnamefont{Bruns}} \bibnamefont{and}
	\bibinfo{author}{\bibfnamefont{D.~A.} \bibnamefont{Tortorelli}},
	\bibinfo{journal}{Methods in Applied Mechanics and Engineering}
	\textbf{\bibinfo{volume}{190}}, \bibinfo{pages}{3443} (\bibinfo{year}{2001}).
	
	\bibitem[{\citenamefont{Sigmund}(1997)}]{Sigmund:97}
	\bibinfo{author}{\bibfnamefont{O.}~\bibnamefont{Sigmund}},
	\bibinfo{journal}{Mechanics of Structures and Machines}
	\textbf{\bibinfo{volume}{25}}, \bibinfo{pages}{493} (\bibinfo{year}{1997}).
	
	\bibitem[{\citenamefont{Allaire and Kohn}(1993)}]{Allaire:93}
	\bibinfo{author}{\bibfnamefont{G.}~\bibnamefont{Allaire}} \bibnamefont{and}
	\bibinfo{author}{\bibfnamefont{R.~V.} \bibnamefont{Kohn}}, in
	\emph{\bibinfo{booktitle}{Topology design of structures}}, edited by
	\bibinfo{editor}{\bibfnamefont{M.~P.} \bibnamefont{Bendsøe}}
	\bibnamefont{and} \bibinfo{editor}{\bibfnamefont{C.~A.~M.}
		\bibnamefont{Soares}} (\bibinfo{publisher}{Kluwer Academic Publisher},
	\bibinfo{year}{1993}), pp. \bibinfo{pages}{207--218}.
	
	\bibitem[{\citenamefont{Jensen and Sigmund}(2005)}]{Sigmund:05}
	\bibinfo{author}{\bibfnamefont{J.~S.} \bibnamefont{Jensen}} \bibnamefont{and}
	\bibinfo{author}{\bibfnamefont{O.}~\bibnamefont{Sigmund}},
	\bibinfo{journal}{Journal of Optical Society of America B}
	\textbf{\bibinfo{volume}{22}}, \bibinfo{pages}{1191} (\bibinfo{year}{2005}).
	
	\bibitem[{\citenamefont{Sigmund}(2007)}]{Sigmund:07}
	\bibinfo{author}{\bibfnamefont{O.}~\bibnamefont{Sigmund}},
	\bibinfo{journal}{Structural and Multidisplinary Optimization}
	\textbf{\bibinfo{volume}{33}}, \bibinfo{pages}{401} (\bibinfo{year}{2007}).
	
	\bibitem[{\citenamefont{Shin and Fan}(2012)}]{Shin:12}
	\bibinfo{author}{\bibfnamefont{W.}~\bibnamefont{Shin}} \bibnamefont{and}
	\bibinfo{author}{\bibfnamefont{S.}~\bibnamefont{Fan}},
	\bibinfo{journal}{Journal for Computational Physics}
	\textbf{\bibinfo{volume}{231}}, \bibinfo{pages}{3406} (\bibinfo{year}{2012}).
	
	\bibitem[{\citenamefont{Shin and Fan}(2013)}]{Shin:13}
	\bibinfo{author}{\bibfnamefont{W.}~\bibnamefont{Shin}} \bibnamefont{and}
	\bibinfo{author}{\bibfnamefont{S.}~\bibnamefont{Fan}},
	\bibinfo{journal}{Optics Express} \textbf{\bibinfo{volume}{21}},
	\bibinfo{pages}{22578} (\bibinfo{year}{2013}).
	
\end{thebibliography}
\end{document}